# MECHANISM OF CANTED MAGNETIC STRUCTURE FORMATION IN THE ABSENCE OF SPIN-ORBITAL INTERACTION

E.V. Rosenfeld

Institute of Metal Physics, Ural Division, Russian Academy of Sciences, 620041, Ekaterinburg

A simple exactly solvable model of canted magnetic structure appearance in the system of crystallographic and chemically equivalent atoms is proposed. The corresponding mechanism originates from the competition of intra- and interatomic exchange interactions.

PACS nomber: 75.45.+j   Macroscopic quantum phenomena in magnetic systems

Ternary rare-earth compounds $RT_2X_2$ (R - rare-earth, T – transition metal, X – Si, Ge) with the tetragonal $ThCr_2Si_2$-type structure (space group I4/mmm) демонстрируют целый ряд необычных физических свойств [1,2]. Особое место в этом ряду занимают изображенные на рис. 1 магнитные структуры, возникающие здесь при T =Mn и частичном замещении R=La другим редкоземельным элементом (Sm [2] или Y [3]). Магнитные атомы Mn в этих соединениях упакованы в слои с плоской квадратной решеткой, и внутри каждого слоя перпендикулярные его плоскости составляющие магнитных моментов упорядочены ферро-, а параллельные – антиферромагнитно (FM and AFM correspondingly). Суммарные магнитные моменты слоев также упорядочиваются либо ферро-, либо антиферромагнитно (см. рис. 1), и при изменении давления [2] или температуры [3] происходит скачкообразный переход между этими типами упорядочения, причем величина моментов слоев почти не меняется. Это, очевидно, означает, что межслойное обменное взаимодействие играет второстепенную роль, и образование скошенной магнитной структуры (CMS) обусловлено взаимодействиями внутри слоя, все атомы которого кристаллографически и химически эквивалентны.

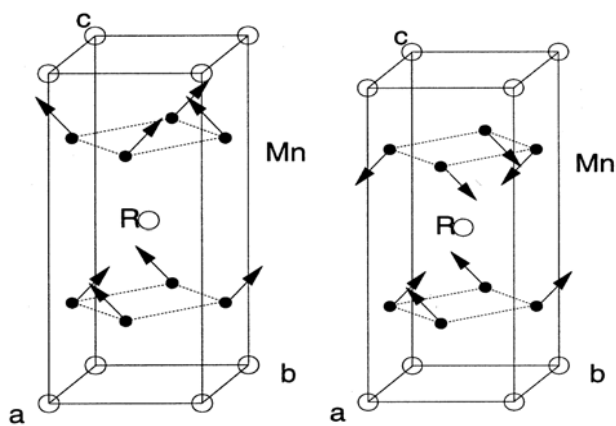

**Рисунок 1**. Возможные типы магнитного упорядочения в $RMn_2Si_2$ [3].

Такая картина очень похожа на наблюдаемую в слабых ферромагнетиках, но микроскопический механизм образования CMS в $RT_2X_2$ должен быть совершенно иным уже просто потому, что при этой кристаллической структуре оба предложенных Morya [4] механизма скашивания не работают. Более того, угол скашивания велик (порядка 45 - 60°), и с ростом температуры ферромагнитная составляющая внутри слоя исчезает *раньше*, чем антиферромагнитная, $T_C < T_N$ [2].

Не выходя за рамки обычной модели Гейзенберга, вряд ли можно объяснить возникновение в решетке Браве FM упорядочения одних составляющих момента и AFM - других. В этой модели полный спин атома **S** является хорошим квантовым числом, т.е. хундовское обменное взаимодействие *I*, по порядку величины близкое к 1 eV, считается бесконечно сильным. Поэтому обменное взаимодействие $\hat{H}_{exch} = -J_{ab}(\hat{\mathbf{S}}_a \cdot \hat{\mathbf{S}}_b)$ пары атомов *a* и *b* с полным спином $\hat{\mathscr{S}} = \hat{\mathbf{S}}_a + \hat{\mathbf{S}}_b$ описывается единственным обменным интегралом $J_{ab}$, и только его знак определяет тип упорядочения. При $J_{ab} > 0$ основным для такой пары является состояние с $\mathscr{S} = 2S$, и в кристалле возникает FM упорядочение; при $J_{ab} < 0$ в основном состоянии $\mathscr{S} = 0$, и устойчива AFM-structure. Прямые численные расчеты, начиная с работ A. Freeman and R. Watson [5] показывают, однако, что обменные интегралы между разного типа 3*d*-orbitals соседних узлов могут достигать вполне сравнимой с *I* величины порядка 0.1 eV и иметь *разные знаки*.

Ниже показано, что при учете конкуренции между внутри- и межатомными обменными взаимодействиями электронов на разных орбиталях суммарный спин пары атомов в основном состоянии может иметь промежуточные значения $0 < \mathscr{S} < 2S$, что ведет к возникновению в кристалле CMS.

Пусть на атоме *a* в двух орбитальных состояниях 1 и 2 находятся два неспаренных электрона со спинами $\hat{\mathbf{s}}_1$ и $\hat{\mathbf{s}}_2$, связанных обменным взаимодействием *I*. Соответствующие состояния на соседнем таком же атоме *b* будем обозначать индексами 3 и 4 и предположим, что обменные интегралы между парами эквивалентных состояний 1,3 ($J_1$) и 2,4 ($J_2$) имеют разные знаки, а всеми остальными взаимодействиями можно пренебречь. Гамильтониан для такой пары атомов будет иметь очень простой вид

$$\hat{H}_0 = -I\{(\hat{\mathbf{s}}_1 \cdot \hat{\mathbf{s}}_2) + (\hat{\mathbf{s}}_3 \cdot \hat{\mathbf{s}}_4)\} + J_1(\hat{\mathbf{s}}_1 \cdot \hat{\mathbf{s}}_3) - J_2(\hat{\mathbf{s}}_2 \cdot \hat{\mathbf{s}}_4); \quad I, J_1, J_2 > 0, \qquad (1)$$

Прежде чем переходить к исследованию этой квантовой системы, обсудим кратко поведение ее классического аналога. Полагая $\mathbf{s}_1, ... \mathbf{s}_4$ обычными векторами и выбирая оси координат, как показано на рис. 2, для энергии получим выражение

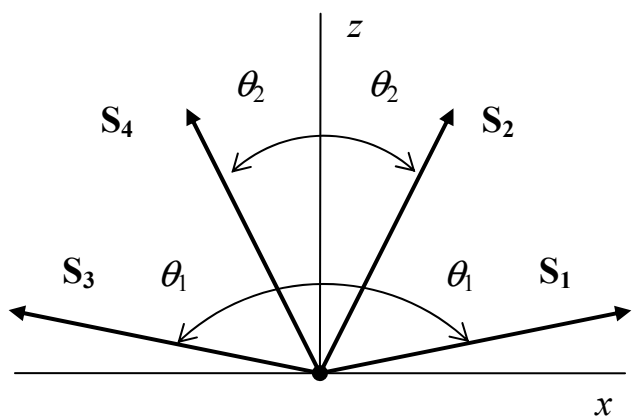

**Рисунок 2**. Система четырех классических векторов с энергией (2).

$$E = -2I\cos(\theta_1 - \theta_2) + J_1 \cos(2\theta_1) - J_2 \cos(2\theta_2). \qquad (2)$$

Такая система с антиферромагнитным обменным взаимодействием между $s_1$ и $s_3$ и ферромагнитным - между $s_2$ и $s_4$ в точности эквивалентна системе, в которой ось $z$ является «трудной» для $s_1$ и $s_3$ и «легкой» – для $s_2$ и $s_4$ (константы анизотропии $2J_1$ и $-2J_2$ соответственно). В последнем случае мы приходим к модели двухподрешеточного магнетика со взаимно перпендикулярными осями анизотропии подрешеток [6], очень схожей со вторым из предложенных Morya [4] механизмов возникновения скашивания.

При обсуждении (2) удобно ввести углы между спинами внутри одного атома $\delta = \theta_1 - \theta_2$ и между спинами атомов $S_a$ и $S_b$ $2\theta = \theta_1 + \theta_2$. Если хундовское взаимодействие $I \to \infty$, спины внутри атомов параллельны ($\delta \to 0$), угловая зависимость энергии (2) принимает вид $(J_1 - J_2)\cos(2\theta)$, и $S_a$ и $S_b$ параллельны друг другу и оси $z$ при $J_1 - J_2 < 0$ или направлены навстречу друг другу вдоль оси $x$ при $J_1 - J_2 > 0$. Однако, для достаточно близких по величине $J_1$ и $J_2$

$$-1 < \frac{I \cdot (J_1 - J_2)}{2 J_1 J_2} < 1 \qquad (3)$$

при любых сколь угодно больших значениях $I$ возникает неколлинеарная структура

$$\tan(\delta) = \frac{(J_1 + J_2)\sin(2\theta)}{2I + (J_2 - J_1)\cos(2\theta)},$$
$$\cos(2\theta) = \frac{I \cdot (J_2 - J_1)}{2 J_1 J}. \qquad (4)$$

При переходе через правую границу неравенства (3) $\theta_1 = \theta_2 = \pi/2$, через левую - $\theta_1 = \theta_2 = 0$, но если (3) выполнено, появляется конечный угол $\delta$ между $s_1$ и $s_2$, и по мере изменения $J_1$ и $J_2$ идет плавный разворот $\mathbf{S}_a$ и $\mathbf{S}_b$ от ферромагнитного к антиферромагнитному упорядочению.

Все $2^4=16$ собственных состояний гамильтониана (1) можно найти в явном виде. Это два синглета, три триплета и квинтет, кратность соответствует значению полного спина пары атомов $\mathscr{S}$. Девять состояний в нижней части спектра - синглет, триплет и квинтет – имеют, соответственно, энергии

$$E_0 = \frac{2I - J_1 + J_2}{4} - \tilde{K}, \quad E_1 = -\frac{J_1 - J_2}{4} - \frac{1}{2}K, \quad E_2 = \frac{-2I + J_1 - J_2}{4};$$
$$K = \sqrt{I^2 + (J_1 + J_2)^2}, \quad \tilde{K} = \sqrt{I^2 + \tfrac{1}{2}I(J_1 - J_2) + \tfrac{1}{4}(J_1 - J_2)^2}.$$
(5)

Остальные семь состояний имеют энергии примерно на $I$ выше, поскольку не удовлетворяют правилу Хунда – среднее значение квадрата атомного момента в них значительно меньше двойки. В состоянии же с $\mathscr{S} = 2$ это среднее точно равно, а в состояниях (5) с $\mathscr{S} = 1$ и $\mathscr{S} = 0$ - лишь немного меньше двойки:

$$\langle 1 | (\hat{\mathbf{S}})^2 | 1 \rangle = \tfrac{1}{2}(3 + I/K) \approx 2 - (J_1 + J_2)^2 / (2I)^2,$$
$$\langle 0 | (\hat{\mathbf{S}})^2 | 0 \rangle = 1 + (4I + J_1 - J_2)/(4\tilde{K}) \approx 2 - (\tfrac{3}{32})(J_1 - J_2)^2 / I^2.$$
(6)

Полагая $I \sim 1$ eV, $J_{1,2} \sim 0.1$ eV, получаем, что отклонение среднего спина атома от 1 в этих состояниях не превышает 1%. Соответствующий проигрыш в хундовской энергии, связанный с уменьшением корреляторов $\langle (\hat{\mathbf{s}}_1 \hat{\mathbf{s}}_2) \rangle$ и $\langle (\hat{\mathbf{s}}_3 \hat{\mathbf{s}}_4) \rangle$, с избытком компенсируется выигрышем в энергии межатомного обменного взаимодействия из-за уменьшения $\langle (\hat{\mathbf{s}}_1 \hat{\mathbf{s}}_3) \rangle$ и возрастания $\langle (\hat{\mathbf{s}}_2 \hat{\mathbf{s}}_4) \rangle$. Этот выигрыш настолько велик, что при выполнении условия

$$-1 < \frac{I \cdot (J_1 - J_2)}{2 J_1 J_2} < \frac{2I^2 - J_1 J_2}{I^2 - 2 J_1 J_2}$$
(7)

(сравни с (3)) состояние с $\mathscr{S} = 1$ вообще оказывается основным. Повторим, что это в принципе невозможно в пределе $I \to \infty$, в котором мы получаем один межатомный обменный интеграл $J_{ab} = \tfrac{1}{4}(J_1 - J_2) \sim 0.01$ eV.

Чтобы определить магнитную структуру кристалла с решеткой Браве, в котором пары ближайших соседей связаны взаимодействием (1), учтем взаимодействие каждого атома выделенной пары с остальными $n$-1 соседними атомами в приближении молекулярного поля

$$\hat{V} = (n-1)\left\{\hat{V}_d + \hat{V}_{nd} - 0.5\left[J_1(\langle \mathbf{S}_1\rangle \cdot \langle \mathbf{S}_3\rangle) - J_2(\langle \mathbf{S}_2\rangle \cdot \langle \mathbf{S}_4\rangle)\right]\right\},$$
$$\hat{V}_d = J_1\left(\langle S_3^z\rangle \hat{S}_1^z + \langle S_1^z\rangle \hat{S}_3^z\right) - J_2\left(\langle S_4^z\rangle \hat{S}_2^z + \langle S_2^z\rangle \hat{S}_4^z\right), \quad (8)$$
$$\hat{V}_{nd} = J_1\left(\langle S_3^x\rangle \hat{S}_1^x + \langle S_1^x\rangle \hat{S}_3^x\right) - J_2\left(\langle S_4^x\rangle \hat{S}_2^x + \langle S_2^x\rangle \hat{S}_4^x\right).$$

При низких температурах и больших $J_2$ система, очевидно, ферромагнитна: основным для каждого атома является состояние с $S=S^z=1$, а для пары - $\mathscr{S}=\mathscr{S}^z=2$, см. (5). Если при уменьшении $J_2$ какие-либо средние значения $x$-проекций спина окажутся отличными от нуля, non-diagonal часть $V_{nd}$ оператора (8) начинает подмешивать к основному собственные состояния $\hat{H}_0$ с $\mathscr{S}^z=1$. Из рассмотрения при этом следует исключить подмешивание состояний, входящих в квинтет: это дает $\langle S_1^x\rangle = \langle S_3^x\rangle$, $\langle S_2^x\rangle = \langle S_4^x\rangle$ и сводится к тривиальному вращению намагниченности. Если же выполняется условие $\langle S_1^x\rangle = -\langle S_3^x\rangle$, $\langle S_1^x\rangle = -\langle S_3^x\rangle$, т.е. возникает CMS, происходит подмешивание состояний из триплетов, в основном из нижнего, имевшего в отсутствие молекулярного поля энергию $E_1$ (5), и чем ближе $E_1$ к $E_2$, тем сильнее это подмешивание. Используя явные выражения для волновых функций, можно получить точную формулу для соотношения обменных интегралов, при котором возникающее infinitesimal canting не вызывает изменения энергии:

$$\frac{J_2 - J_1}{I} = \frac{1}{n}\left\{\sqrt{1 + \left(n\frac{J_2 + J_1}{I}\right)^2} - 1\right\} \approx \frac{n}{2}\left(\frac{J_2 + J_1}{I}\right)^2. \quad (9)$$

Именно эта формула определяет линию границы между FM и CMS фазами на фазовой диаграмме, изображенной на рис.3.

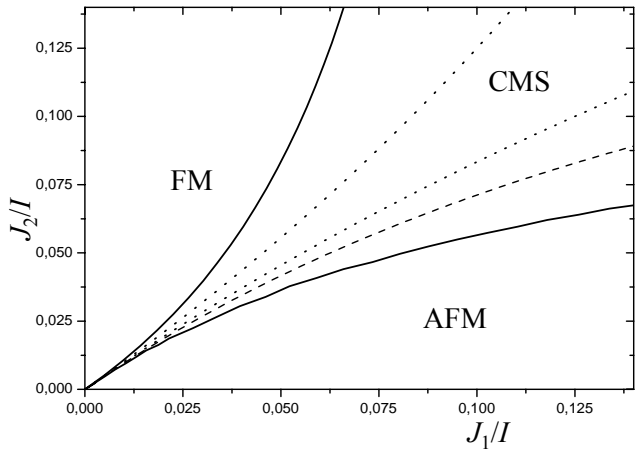

**Рисунок 3**. Фазовая диаграмма при $T=0$. Жирные сплошные линии – границы между FM, AFM и CMS - фазами согласно (9) и численным расчетам, Dot and Dash – границы неравенства (3) и правая граница (7) соответственно.

Чтобы получить аналогичную формулу для линии границы между AFM и CMS фазами, нужно определить основное состояние пары атомов в AFM-окружении, что вряд ли можно сделать в аналитическом виде. Следует заметить, однако, что линия этой границы, полученная в результате численных расчетов (см. рис.3), с хорошей точностью совпадает с линией, получающейся из (9) в результате замены $J_1 \rightleftarrows J_2$. Заметим также, что поскольку направление намагниченности возникающей в этой модели CMS может быть произвольным, такая структура не может описываться инвариантом Дзялошинского [7]. Учитывая (4), для ее описания естественно выбрать энергию в виде

$$E = (J_1 - J_2)(\mathbf{S}_a \cdot \mathbf{S}_b) + \frac{J_1 J_2}{I}(\mathbf{S}_a \cdot \mathbf{S}_b)^2. \qquad (10)$$

Наконец, на рис. 4 показаны результаты самосогласованных вычислений средних значений проекций атомного момента (z-проекции упорядочиваются ферро-, x-проекции - антиферромагнитно). Видно, что если при низких температурах устойчива CMS, переход из нее в парамагнитную (PM) фазу с ростом температуры, в зависимости от соотношения между $J_1$ и $J_2$, идет через FM- или AFM структуру. Для всех трех изображенных на рис. 4 случаев соотношения параметров $J_1 = 1000$, $J_2 = 850$; $J_1 = 500$, $J_2 = 650$; $J_1 = 50$, $J_2 = 200$ K разность $|J_1 - J_2| = 150$ K постоянна, так что в пределе $I \to \infty$ температура перехода в парамагнитную фазу

$$T_{C,N} = \frac{nJS(S+1)}{3} = \frac{n|J_1 - J_2|}{6}, \qquad (11)$$

показанная стрелкой, также постоянна и при $n=4$ равна 100 K. Таким образом, если предлагаемая модель правильно объясняет причину возникновения CMS, кроме последовательности переходов CMS→AFM→PM, наблюдаемой в $RT_2X_2$, можно ожидать обнаружения последовательности CMS→FM→PM.

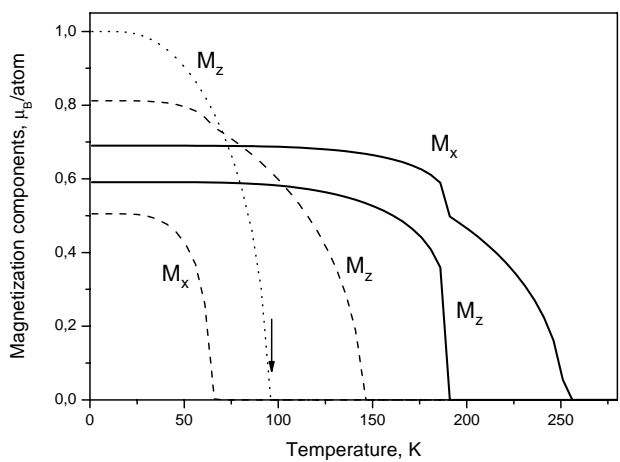

**Рисунок** 4. Температурная зависимость составляющих намагниченности (ферромагнитное упорядочение вдоль z, антиферромагнитное – вдоль x). Сплошные линии - $J_1 = 1000$, $J_2 = 850$ K, Dash - $J_1 = 500$, $J_2 = 650$ K, Dot - $J_1 = 50$, $J_2 = 200$ K. Во всех случаях $I = 10000$ K, $n = 4$, стрелка указывает температуру перехода порядок-беспорядок в пределе $I \to \infty$ (11).

Учитывая эти выводы, можно поставить более общий вопрос о границах применимости простой модели, в которой при описании обменного взаимодействия многоэлектронных атомов используется единственный параметр. Если его величина оказывается существенно меньше, чем характерные для данного межатомного расстояния значения обменных интегралов между отдельными орбиталями, переход к более подробному описанию может привести к значительной перестройке спектра, когда речь идет о молекулах, и к переоценке устойчивости разного типа магнитных структур в кристаллах.